\documentclass[sigconf]{acmart}

\AtBeginDocument{%
  }

\copyrightyear{2025}
\acmYear{2025}
\setcopyright{rightsretained}
\acmConference[]{}{}{}
\acmBooktitle{}
\acmDOI{10.1145/3715675.3715800}

\usepackage[capitalize]{cleveref}
\usepackage{caption}
\usepackage{subcaption}
\usepackage{float}
\usepackage[htt]{hyphenat} 
\usepackage{array}
\usepackage{arydshln}

\usepackage{svg}
\usepackage{algorithm}
\usepackage{algpseudocode}
\usepackage{pifont}
\usepackage{lipsum}  
\usepackage{diagbox}
\usepackage{multirow}
\usepackage[export]{adjustbox}

\crefname{section}{Sec.}{Secs.}
\Crefname{section}{Section}{Sections}
\Crefname{table}{Table}{Tables}
\crefname{table}{Tab.}{Tabs.}

\begin{document}

\title{Scalable Event-Based Video Streaming for Machines with MoQ}

\author{Andrew C. Freeman}
\email{andrew_freeman@baylor.edu}
\orcid{0000-0002-7927-8245}
\affiliation{%
  \institution{Baylor University}
  \city{Waco}
  \state{Texas}
  \country{USA}
}

\renewcommand{\shortauthors}{Freeman}

\begin{abstract}

Lossy compression and rate-adaptive streaming are a mainstay in traditional video steams. However, a new class of neuromorphic ``event'' sensors records video with asynchronous pixel samples rather than image frames. These sensors are designed for computer vision applications, rather than human video consumption. Until now, researchers have focused their efforts primarily on application development, ignoring the crucial problem of data transmission. We survey the landscape of event-based video systems, discuss the technical issues with our recent scalable event streaming work, and propose a new low-latency event streaming format based on the latest additions to the Media Over QUIC protocol draft.

\end{abstract}

\ccsdesc[500]{Information systems~Multimedia streaming}
\ccsdesc[300]{Networks~Application layer protocols}
\ccsdesc[100]{Computing methodologies~Computer vision}

\keywords{SVC, DVS, event camera, streaming, QUIC, MoQ, event-based vision, event video }

\received{15 December 2024}

\maketitle

\section{Introduction}
\label{sec:intro}

After decades of research and development, video streaming has come to constitute a substantial amount of Internet traffic. There has been substantial progress in the underlying video codecs, the adaptation mechanisms, and the streaming protocols. Scalable video coding (SVC) \cite{McCanne:CSD-96-928,schwarz_overview_2007} showed early promise with its enhancement layer mechanism, but decoder overhead has limited its adoption. Adaptive bitrate (ABR) streaming is much more common \cite{yang_opportunities_2014}. Here, the publisher encodes a video at several bitrates, and the client simply requests the stream best suited to its network conditions. Various streaming protocols have arisen to suit different needs, such as HLS for video on demand and WebRTC for teleconferencing.

To date, these streaming systems have focused on traditional frame-based video. In recent years, however, novel \textit{event cameras} have gained traction in the computer vision and robotics communities. These sensors do not capture image frames; rather, each pixel senses asynchronously, outputting a discrete timestamped ``event'' when its log intensity change exceeds a certain threshold \cite{gallego_event-based_2022}. Since intensity change occurs most dramatically at points of high contrast change, a stationary event camera predominantly outputs events for the edges of moving objects. These cameras achieve microsecond temporal resolution, high dynamic range ($>$120 dB), and low power usage \cite{gallego_event-based_2022}, but suffer from extremely high data rates. Event-based computer vision applications have shown compelling performance, but their relative slowness and power usage undermines many of the benefits of the sensor technology.

\begin{table*}
  \centering
  {\small{
  \begin{tabular}{@{}lccccp{0.07\linewidth}p{0.08\linewidth}cp{0.04\linewidth}p{0.07\linewidth}@{}}
    \toprule
    Year & Paper  & Type  & Event repr. & Inference speed (ms) & GPU & GPU Cost & GPU TDP (W) \\
    \midrule
    2021 & Nested-T \cite{zhang_nested_2021,peng_get_2023} & ViT & Attention embedding & 32.1* &  1080 Ti & \$200 & 250 \\ 
    2022 & Swin-T v2 \cite{liu_swin_nodate} & ViT & Attention embedding & 33.6* &   1080 Ti & \$200 & 250 \\
    2022 & ASTMNet \cite{li_asynchronous_2022} & CNN + RNN & Attention embedding & 72.3* & Titan Xp & \$200 & 250 \\
    2023 & GET-T \cite{peng_get_2023} & GET & Attention embedding & 17.8* & 1080 Ti & \$200 & 250 \\
    2023 & DMANet \cite{wang2023dual} & CNN + RNN & EventPillar & 30.2 & 2080 Ti & \$300 & 250 \\
    2023 & ERGO-12 \cite{zubic2023chaos} & ViT + RNN & ERGO-12 & 101.1 & T4 & \$700 & 70 \\ 
    2023 & RVT-B \cite{gehrig_recurrent_2023}  & ViT + RNN & 2D Histogram & 11.9 & T4 & \$700 & 70 \\
    2024 & SAST \cite{peng2024scene} & ViT + RNN & Event Voxel & 19.7 & Titan Xp & \$200 & 250 \\
    2024 & SpikingViT \cite{yu_spikingvit_2024} & ViT + TMSN & 2D Histogram & 26.9 & 3090 & \$900 & 350 \\
    2024 & STAT \cite{guo2024spatio} & ViT + TAM & 2D Histogram & 83.6 & Titan Xp & \$200 & 250 \\ 
    2024 & STF \cite{zhu2024spatio} & CNN + RNN & 2D Histogram & 48.8 & Titan Xp & \$200 & 250 \\
    2024 & S5-ViT-B \cite{zubic2024state} & ViT + SSM & 2D Histogram & 9.6 & T4 & \$700 & 70 \\
    2024 & DTSDNet-M \cite{fan2024dense} & CNN & 2D Histogram & 7.4 & 4090 & \$1600 & 450 \\

    \bottomrule
  \end{tabular}
  }}
  \caption{Speed comparisons on the 1 Mpx dataset. GPU prices are reported in USD based on the ``Buy It Now'' price on eBay at the time of writing (late 2024). A * denotes that event representation construction is included in the inference speed.}
  \label{tab:application_survey}
\end{table*}

We argue that the research emphasis on high power, low rate applications is missing a crucial component for real-world systems: adaptive streaming. In this paper, we survey event-based object detection as an example of a GPU-based application, focusing on the reported speed and energy costs. We then discuss the necessity of event compression and streaming for the practical deployment of these applications, and offer preliminary results from our own investigation into rate-adaptive event streaming. Based on these results and recent developments in the Media Over QUIC (MoQ) Transport draft, we propose a new format for streaming event camera data in a scalable manner with MoQ. This format takes advantage of the data agnosticism of MoQ to achieve low latency and scalable adaptation using the existing protocol mechanics.


\section{Event-Based Video}

\begin{figure*}
  \centering
    \begin{subfigure}{0.33\textwidth}
        \centering
        \includegraphics[width=\linewidth]{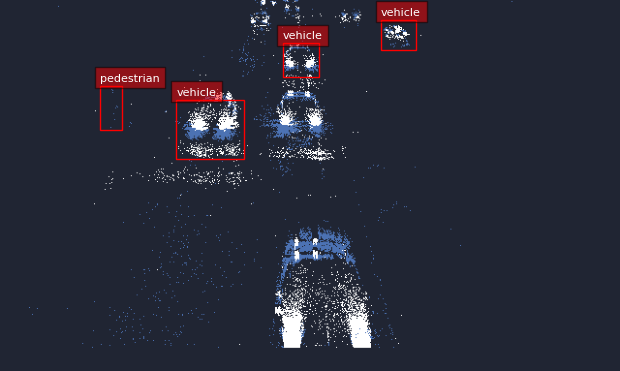}
        \caption{100 Mbps}
        \label{fig:teaser_100}
    \end{subfigure}
    \begin{subfigure}{0.33\textwidth}
        \centering
        \includegraphics[width=\linewidth]{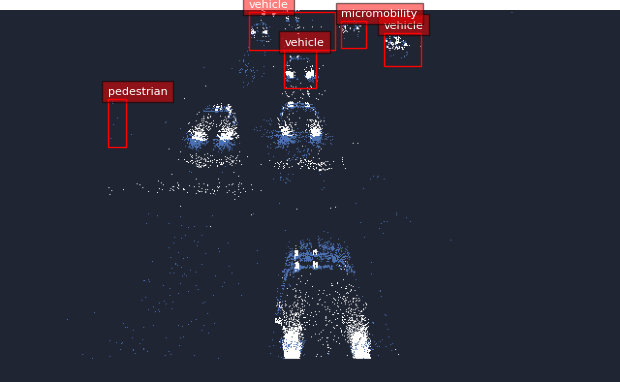}
        \caption{50 Mbps}
    \end{subfigure}
    \begin{subfigure}{0.33\textwidth}
        \centering
        \includegraphics[width=\linewidth]{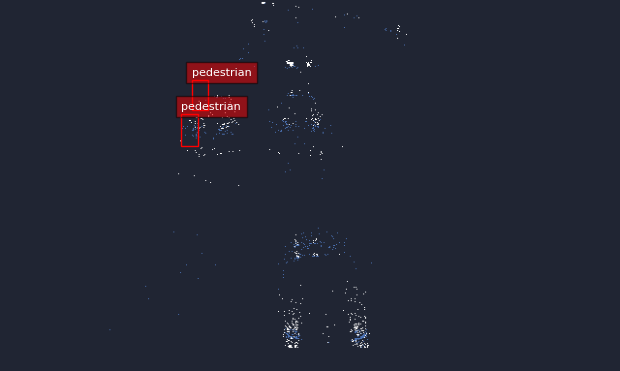}
        \caption{5 Mbps}
    \end{subfigure}
    \caption{Examples of event reduction at various bandwidth limits, with object detections overlaid (repeated from our prior work \cite{hamara_low-latency_2024}). Events were accumulated over a period of 50 ms to generate the framed representations. Bright blue pixels indicate negative-polarity events, while white pixels indicate positive-polarity events. Dark blue background pixels indicate the absence of any event. At lower bandwidths, there are fewer events in the raw representation, lowering the application accuracy.}
    \label{fig:five_images}
\end{figure*}

A pixel in an event camera continuously measures the log incident intensity \cite{gallego_event-based_2022}. When the log intensity changes beyond a given threshold (gets brighter or darker by a certain amount), the camera outputs a tuple of the form $\langle x, y, t, p \rangle$ \cite{gallego_event-based_2022}. Here, $x$ and $y$ are the spatial coordinates, $t$ is a microsecond-resolution timestamp, indicating the precise moment that the change threshold was met, and $p$ is the 1-bit change polarity. We offer example visualizations of event camera data in \cref{fig:five_images}.

Where traditional cameras have a fixed data rate in the uncompressed representation, the raw data rate for event cameras depends entirely on the amount of motion being recorded. Under high motion, a 720p event camera can easily produce raw data at a rate exceeding 500 Mbps, compared to a fixed rate of 221 Mbps for a monochrome framed camera at 30 FPS. The microsecond temporal precision and 32-bit timestamps of a typical event camera gives it a high-speed view of the world similar to that of a 1-million FPS framed camera, at a fraction of the raw data rate, weight, and power consumption. Since the data is spatiotemporally sparse, and it does not express absolute intensities, however, these cameras are designed for computer vision applications, rather than human viewership.

\subsection{The Limitations of Event-Based Vision with GPUs}
With the high data rate demands of event cameras, many computer vision applications require significant computational and energy resources to operate. Although specialized neuromorphic application hardware can dramatically increase efficiency, GPUs are much more common in the literature due to their accessibility and support infrastructure. For GPU-based applications, the event streams must be converted to a variety of frame-based representations, such as event polarity \cite{rebecq2017real, moeys2016steering} and event count \cite{Maqueda_2018_CVPR} histograms,  Surface of Active Events (SAE)\cite{park2016performance, benosman2013event, zhu2018ev}, voxel grids \cite{zhu2019unsupervised, peng2024scene}, and Event Pillars \cite{wang2023dual}. To maintain ``real-time'' performance, the application will temporally group the events into framed representations at a frequency determined by the inference speed. Many works that merely operate at a rate of 20-100 Hz then claim to be ``real-time.'' While such a claim is reasonable with a frame-based camera, it is less convincing when an event camera records with 1 million Hz precision. In the literature, if events are streamed (e.g., from a small robot to a more powerful machine), the devices are on the same network and the distance is extremely short. Some efforts have explored lossless compression of event data \cite{bi_spike_2018,khan_lossless_2020,schiopu_entropy_2023}; however, many event-based vision applications can maintain high accuracy when many events are discarded entirely \cite{gruel_performance_2023,freeman_rethinking_2024}.

A heavily-explored application for event-based vision is object detection. Numerous datasets provide benchmarks of comparison between different methods. Examining one such dataset, the 1 Mpx automotive detection dataset, we provide the author-reported speed benchmarks from the literature in \cref{tab:application_survey}.

We see that the fastest reported inference speed is 7.4, equating to an operating rate of 135 Hz. In practice, the realized speed is slower, since there is addtional latency in constructing the event representation and transferring the data into the GPU. This speed is a limitation of the model architecture, rather than the sensor modality, meaning that the high-speed capture of the event sensor is not fully utilized. Consumer frame-based cameras, meanwhile, can easily achieve capture rates of 240 FPS. One may argue that the event camera data will avoid the motion blur associated with traditional cameras. However, a framed camera typically has a maximum shutter speed of 1/4000th of a second, avoiding motion blur for all but the fastest-moving objects. 
 
One may then argue that the superior dynamic range and effective ``night vision'' of an event camera set it apart. Framed cameras can capture HDR video through simple exposure stacking, however, and many vision-oriented cameras include infrared (IR) sensors for night recording.

Finally, one may argue that the low power consumption of an event camera allows it to be deployed on small robotic vehicles and edge devices, as the maximum power consumption of the Prophesee camera used in the 1 Mpx dataset capture is only 205 mW \cite{noauthor_event-based_2024}. Meanwhile, the minimum thermal design power (TDP) of the reported GPUs used for inference in \cref{tab:application_survey} is 70 W. An all-in-one device for sensing and processing, then, must still have a substantial power supply and size, undermining the efficiency and compactness of the sensor.

\subsection{The Need for Event Streaming}

Given these caveats, the most compelling use case today for an event-based GPU application is to mount the event camera aboard a small robot, such as a drone, and perform the application processing offboard. The event camera will have lower power and weight requirements than a traditional camera, allowing for a smaller robot or longer deployment. We now face the question: \textit{how does the event stream get from the robot to the GPU server?}

Suppose, for the sake of argument, that our GPU can perform object detection inference in 5 ms. We want to operate it at 200 Hz to better utilize the event camera's high temporal resolution. Based on \cref{tab:application_survey}, one can reasonably expect such an inference speed to be possible within the coming years. This creates several significant challenges for data transmission. 

Since the GPU performs object detection on a frame-based representation, the first intuitive approach might be to cast the events into the framed representation and encode it with standard video codecs (e.g., H.264, H.265) for streaming. However, this approach is fundamentally limited by the maximum encoding speed on low-power hardware. There is significant additional overhead in the transcoding and packaging of videos for live streaming, which is necessary to accommodate changes in network bandwidth.

To perform high-rate application inference, therefore, we must consider streaming the event stream itself. However, this presents its own challenge, as event cameras can generate millions of events per second during high-motion scenarios. For a high-resolution camera, each raw event typically requires 8-16 bytes to encode. Even at low activity levels of 100K events per second, this produces a raw data rate of 6.4-12.8 Mbps. During high motion activity, this event rate would accordingly scale, likely overwhelming the connection to a GPU server. Standard lossless compression techniques (e.g., gzip, LZ4, xz, etc.) offer some relief but do not attain compelling compression ratios or speeds. More importantly, lossless compression cannot adapt to changing network conditions. When bandwidth becomes constrained, the system has no way to gracefully degrade performance.

Therefore, we argue that practical event-based vision systems require lossy, rate-adaptive compression and streaming algorithms specifically designed for event data. Such a system should:

\begin{itemize}
    \item Preserve the microsecond temporal precision of the events and spatial resolution of the sensor
    \item Dynamically adjust compression rates based on network conditions and application latency requirements
    \item Scale efficiently with the camera resolution, event rate, and network configurations
\end{itemize}

Below, we describe the existing work in these areas and provide preliminary results from our end-to-end system for event-based streaming.

\subsection{Event Compression}\label{sec:compression}

Existing techniques for event compression are frequently lossless. The event camera manufacturers each have a proprietary compressed data format, such as AEDAT for iniVation sensors \cite{noauthor_aedat_2024} and EVT for Prophesee sensors \cite{noauthor_evt_2024}. These lightweight formats are designed for on-camera compression to prevent bottlenecking over a USB connection to a computer. Schiopu et al. further introduced two bespoke lossless compression algorithms, substantially outperforming generic encoders such as LZMA \cite{schiopu_low-complexity_2022,schiopu_entropy_2023}. Gruel et al. explored spatial and temporal downscaling for action recognition applications \cite{gruel_performance_2023,gruel_frugal_2023}. Recent work leverages a point cloud representation and the MPEG G-PCC codec, but temporal or spatial quantization is necessary to achieve reasonable speeds \cite{khan_time-aggregation-based_2021,huang_evaluation_2023}. Freeman et al. introduced an event-based, real-time lossy compression system that preserves both the temporal and spatial resolution of the camera, but it requires absolute intensity information derived from a complementary frame-based camera \cite{freeman_asynchronous_2023,freeman_accelerated_2024,freeman_rethinking_2024}. More recently, the Joint Photographic Experts Group (JPEG) has launched an initiative to develop a compression standard for event camera data in JPEG XE \cite{noauthor_jpeg_2024}, but this effort is focused merely on lossless compression for offline or onboard processing.

A straightforward method for inducing loss without quantization is to simply discard some subset of the event stream. Fischer and Milford demonstrated robot localization on a small spatial subset of event sequences \cite{fischer_how_2022}. Banerjee et al. proposed an online compression system for discarding the events outside the regions of interest \cite{banerjee_lossy_2020}, but this method requires a frame-based sensor to establish the regions and determine their priority.

In the context of object detection, we have found that models are resilient to dramatic reductions in the raw event stream. Using the Recurrent Vision Transformer (RVT) model \cite{gehrig_recurrent_2023} trained on the eTraM dataset \cite{verma_etram_2024}, we analyzed the effect of randomized event loss on object detection accuracy. We found that at a bandwidth of 25 Mbps, with 64.9\% of events discarded, the mean average precision (mAP) was reduced by only 0.17 \cite{hamara_low-latency_2024}. Meanwhile, at a 50 Mbps bandwidth and 40.9\% data loss, the mAP was reduced by only 0.13 \cite{hamara_low-latency_2024}. \cref{fig:five_images} shows qualitative examples of one such video at various bitrates, demonstrating the decrease in detection efficacy as bandwidth decreases. Thus, this method provides an adaptive way to reduce data rates without introducing additional computational overhead. Furthermore, the loss does not involve temporal or spatial quantization, leaving room for the application to perform inference at arbitrary rates.

\section{Streaming Protocol}\label{sec:protocol}
In traditional video systems, receiver-driven rate adaptation is critical to enable practical streaming. Methods include adaptive bitrate (ABR), sending various quality versions across different streams, and scalable video coding (SVC) \cite{mccanne_scalable_nodate,schwarz_overview_2007}, sending quality enhancement layers across different streams. In these systems, the receiver of the video can select which stream(s) it wants to receive, according to the current network conditions, processing load, and application-level latency requirements.

To date, there has been no equivalent system for event-based streaming. Since we observed that random event dropping offers a lightweight and effective compression method, we sought to couple the technique with rate adaptation mechanisms to intelligently balance event loss with network conditions.

\subsection{Multi-Track Event Partitioning}\label{sec:old_method}
In our prior work, we used a fork of the work-in-progress Media Over QUIC (MoQ) protocol \cite{gurel_media-over-quic_2024} to construct a number of stream ``tracks'' at the sender \cite{hamara_low-latency_2024}. Specifically, we used \texttt{moq-transfork} version 0.2.0 (commit ID 34a6177) and its associated Rust packages. While we achieved similar results with the pre-fork \texttt{moq-rs} packages, we used \texttt{moq-transfork} for the final evaluation due to authentication issues encountered in the former relay system at the time.

\subsubsection{Prior Results}
To construct our stream tracks, we simply partitioned the event stream by sending an MoQ object of up to $E$ consecutive events on each track. We performed this action at fixed time intervals of 50 ms (matching the inference window of RVT), creating natural join points in the stream. The receiver then subscribed and unsubscribed from tracks as needed to maintain a given latency target. We evaluated the system on a subset of the eTraM test dataset with RVT, using the pre-trained weights provided by the eTraM authors. With a strict end-to-end latency target of 5 ms, we can send only $5000$ events per track, per second, while maintaining our latency target.  With $N=5$ tracks and a network bandwidth of 50 Mbps, this system maintains a mean latency of 2.8 ms with a reduction in mAP of 0.41 \cite{hamara_low-latency_2024}. At a relaxed target latency of 50 ms, we can send up to $50000$ events per track, per second, significantly increasing the throughput. Here, the system maintains a mean latency of 43.2 ms with a 0.24 reduction in mAP \cite{hamara_low-latency_2024}.

\subsubsection{MoQ Synchronization}

This study revealed that time-based event partitioning is a reasonable mechanism to enable rate adaptation. However, our approach to using multiple tracks with simultaneous subscriptions showed a number of limitations. Chiefly, scalability is severely limited by the overhead of concurrent transport streams. During bandwidth-limited scenarios, it was common for the tracks to become desynchronized from one another. That is, a lower-priority track could be 50 ms behind a higher-priority track at the relay, as the relay delays the sending of the lower-priority data. The client is not made aware of the internal relay processes, however, and awaits the arrival of the delayed objects, increasing the end-to-end latency. This problem was exacerbated by the frequent subscription changes (often dozens per second) and the extreme variation in the source data rate (based on the amount of motion in the scene). Although we could detect this desynchronization at the client and unsubscribe from an affected track, there was no mechanism in MoQ for the client to reset the internal relay state when we resubscribe to the track. Hence, the track delay may persist.  This is a known attribute of the current MoQ Transport draft, as documented in issue \#475 \cite{noauthor_no_2024}. The contributors there propose a timestamp-based synchronization mechanism for multiple tracks, but discussion at the time of writing has not arrived at a consensus.

\subsubsection{Inflexibility}

\begin{figure}
  \centering
    \includegraphics[width=\linewidth]{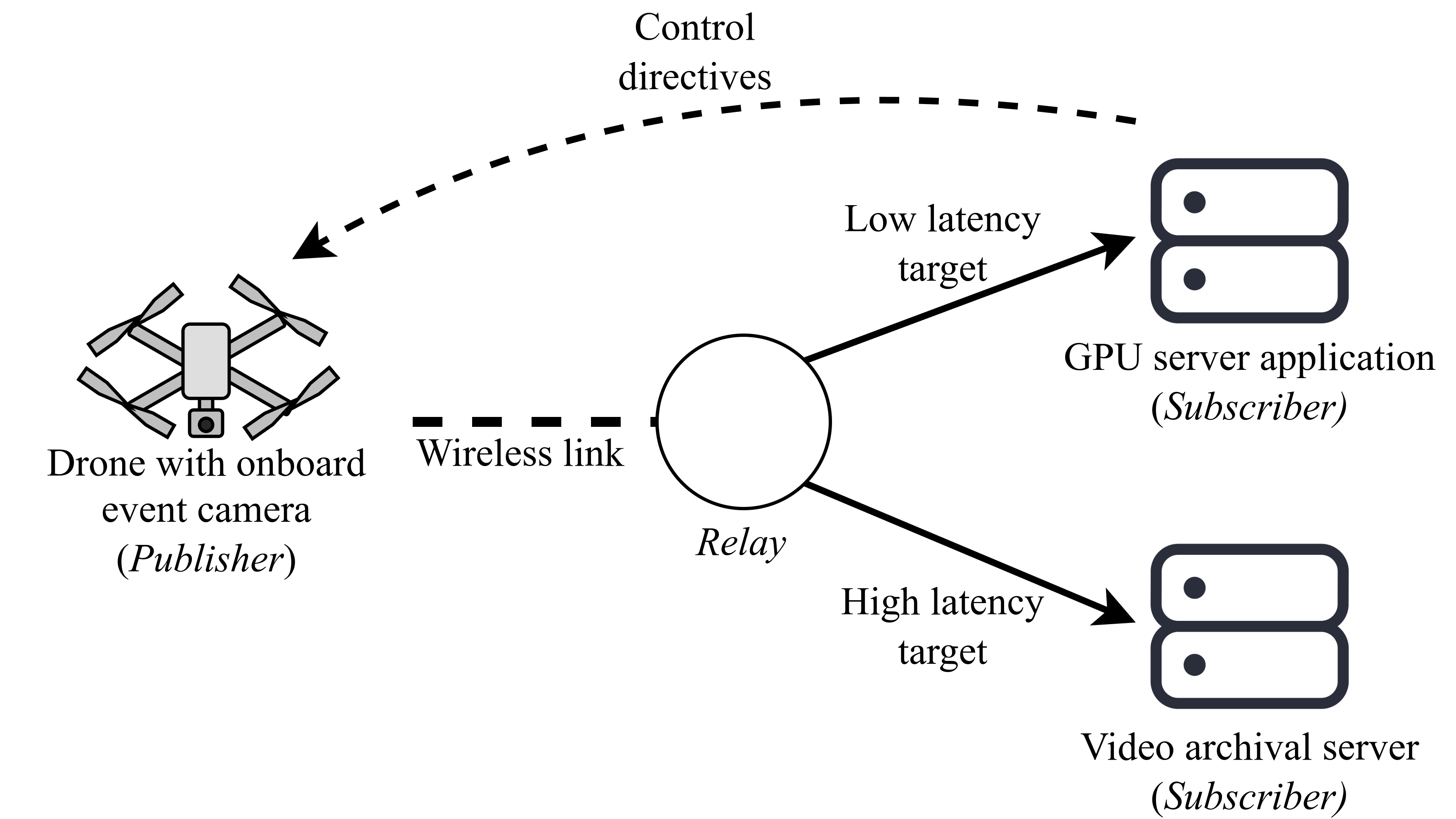}
    \caption{Overview of the heterogeneous applications enabled by our proposed system. A low-latency receiver can set a short delivery timeout for its MoQ subscription, receiving a subset of the event data at high speed. Vision application results may then be used to send control directives back to the drone device, to perform operations such as obstacle avoidance. At the same time, another receiver can receive all of the available data for archival purposes, albeit with higher latency, by simply setting a high delivery timeout for its subscription. }
    \label{fig:system}
\end{figure}

Furthermore, our prior system implicitly assumed that the publisher device is aware of the minimum target latency. The choice of $E$ (how many events to send per track, per unit time) directly corresponds to the minimum latency achievable by a client. For example, we can maintain 5 ms when $E = 250$, but only 50 ms when $E = 2500$. With a small, fixed number of tracks, $N = 5$, this further limits the maximum quality of the received event stream. This inhibits our higher-level goal of supporting heterogeneous applications. For example, \cref{fig:system} illustrates a drone device sending its event camera data to two subscribers through an MoQ relay. A GPU server has a low latency target, so it is tolerant to heavy data loss. It computes an application result, such as object detection, and sends flight control directives to the drone (possibly via the same, bidirectional MoQ Transport connection). At the same time, we want to store as much data as possible from the camera on a secure archival server, irrespective of the latency incurred. With a fixed budget for $N$ and $E$, we cannot satisfy these diverse latency goals simultaneously: if low latency is necessary, the publisher cannot send all of the data from the camera. This server-side adaptation hurts our ability to scale such a system to multiple receivers with their own application-level goals.

\begin{figure*}
  \centering
    \includegraphics[width=\linewidth]{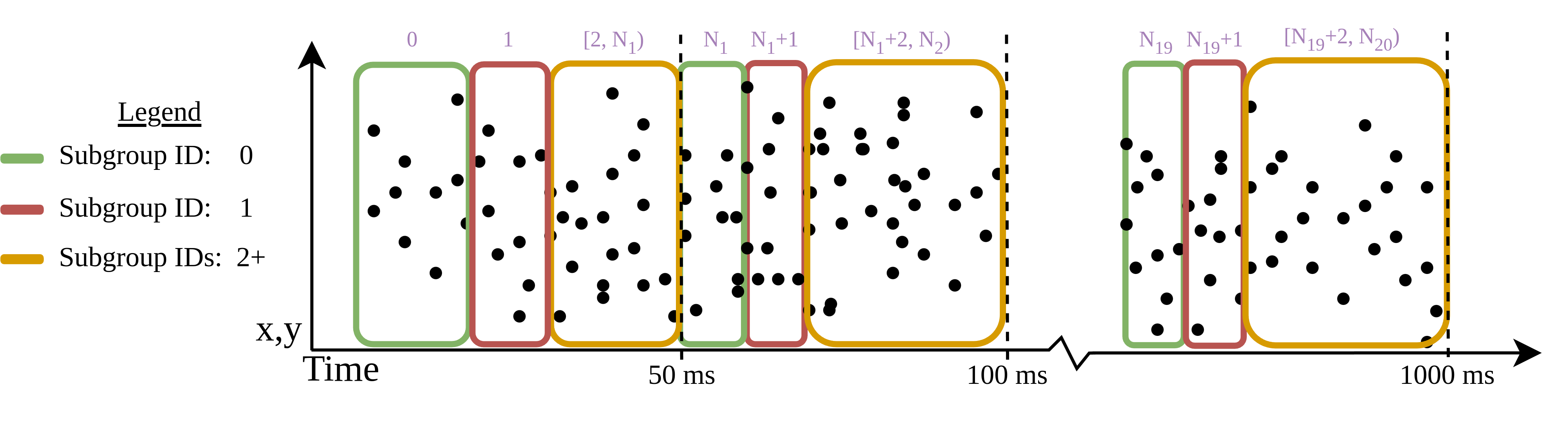}
    \caption{Example of how an event stream may be partitioned into subgroups for MoQ transmission. For simplicity, each dot represents a distinct event in time (x-axis) and space (y-axis), and we do not visualize the polarity. Only the temporal component determines which subgroup an event is placed in. The numbers above each box refer to the object IDs. }
    \label{fig:subgroups}
\end{figure*}

\subsection{Subgroup-Based Event Partitioning}

Meanwhile, the authors of MoQ Transport recently introduced the concept of \textit{subgroups} in Draft 06 \cite{curley_media_2024}. Subgroups add additional granularity to the structure of MoQ data streams. Any group may optionally have a number of subgroups. A relay will attempt to deliver the data from these subgroups according to the underlying subgroup priority and object IDs. With this system, one can ensure that independent data streams remain in temporal synchronization within a single track subscription. 

Draft 06 additionally introduced a delivery timeout mechanism \cite{curley_media_2024}. Here, the subscriber sets the maximum duration that the relay may spend attempting to forward an object. If the timeout duration is reached and the object has not been sent successfully, it is silently dropped for that subscription.

\subsubsection{Design}
With these two constructs, we propose a new approach for partitioning event camera data for scalable streaming. Rather than sending the events across separate tracks, we can partition them across several subgroups in a single track. Each object holds a fixed number of events, $E$, and these objects are placed into subgroups according to their temporal order. Subgroups may hold a variable number of objects per unit time, but each higher-level subgroup should have at least as many objects as the preceding subgroup. Every $T$ milliseconds, the subgroup ID resets to 0. The priority is determined by the subgroup ID, with subgroup 0 having the highest priority.

\cref{fig:subgroups} demonstrates this partitioning scheme. For illustrative purposes, we have $E= 10$ events per object. The object IDs are written across the top of the figure. Subgroups 0 and 1 in this example each have one object per $T = 50$ ms time window. We require a variable number of subgroups depending on the overall number of events per $T$ window. 

By setting a certain delivery timeout for the subscription, we ensure that the relay does not block the progression of any subgroup stream. Since we no longer determine the received data rate by subscribing and unsubscribing to tracks, the delivery timeout becomes our primary adaptation mechanism. If we assume that the objects arrive at the relay instantaneously when they are generated by the camera, then a fixed delivery timeout can ensure that the end-to-end latency is within our target. However, if the connection between the publisher and the relay experiences congestion, we can respond to the received latency by decreasing the delivery timeout for our subscription. Whereas our prior method (\cref{sec:old_method}) frequently required dozens of subscription control messages per second to adapt to the changes in \textit{both} the camera data rate and the overall network bandwidth, this approach frees the client to adapt only to changes in the network bandwidth. In the typical case, we expect to see only a few delivery timeout messages per minute.

While supporting low-latency applications, a separate, higher-latency subscriber, such as the archival server in \cref{fig:system}, may initiate its connection with a very high delivery timeout. This subscriber will then have far fewer objects dropped by the relay. With this system, we can easily experiment with various subgroup event rates \textit{without changing the receiver-driven adaptation algorithm}. For example, we may increase the number of objects per subgroup, per unit time, by a factor of 2 for each increment of the subgroup ID. Then, lower-priority subgroups will carry more event data, but will be less likely to successfully send all their data when there are periods of congestion.

This mechanism does not extend to systems with multiple relays, however, as delivery timeouts are not propagated by relays from the subscribers to the publishers. One possible solution is that relays could communicate their own delivery timeouts to each new subscriber. This information could be cumulative, such that the subscriber is informed of the maximum duration for an object to propagate through all the relays. If this duration is less than the application's target latency, the subscriber can simply set its delivery timeout to the difference between the target latency and the cumulative relay timeouts. This technique would still be limited if early relays have short timeouts, however. This could be somewhat mitigated if a relay opens multiple subscriptions to the \textit{same} published track, but with different delivery timeouts. Careful connection management could avoid sending duplicate objects, and make objects received in a lower-latency subscription available to the higher-latency subscription. In any case, such efforts require further discussion and additions to the MoQ Transport draft.

\subsubsection{Compression Support}

 Variably sized subgroups will be useful as we anticipate future compression schemes for event data. We can achieve a more accurate probability model, and thus higher compression ratios, when events are closely located in both space and time. Thus, we will benefit from longer, continuous segments of events in the same compression context. Therefore, lower-priority subgroups can be expected to achieve better compression characteristics. If an object is dropped, we note that subsequent objects in the subgroup cannot be decoded. The publisher's choice of group interval determines the maximum duration that the receiver may wait before the decoder context is reset for a given subgroup ID. 

\subsubsection{Quality}

We emphasize that this single-track adaptation sch\-eme is designed to maintain extremely low latency, and it is likely ill-suited for traditional video data types. Chiefly, we need not worry about thrashing in the data rate or received video quality. Classical streaming solutions seek to maximize the human quality of experience (QoE) by gradually adjusting the visual quality over time. Event cameras are designed for computer vision applications, however, and event video is largely inscrutable for human viewers even when there is no data loss. We seek only to maximize the application-level performance at a given latency target.

\section{Why MoQ?}
Some may argue that a custom Real-time Transport Protocol (RTP) can yield lower latency than MoQ. Although this may be true, we emphasize that our primary interest at this early stage is to explore receiver-driven adaptation mechanisms for event streaming. The exact latency measurements are unimportant, so long as the application-level performance can map to the latency (and loss) in a predictable manner. In the future, if we require lower latency than MoQ can achieve, we may develop a bespoke RTP-based protocol (e.g., with RTP Over QUIC) with similar adaptation mechanisms.

Meanwhile, event-based vision systems remain a relatively small niche in the research world, and the media-centric mechanisms of MoQ lend themselves well to researchers coming from a vision-oriented background. One can quickly develop prototypes with the existing MoQ implementations, without first having to learn the many components of a monolithic stack such as WebRTC.

Finally, event cameras are often paired with traditional frame-based cameras. Framed sensors provide complementary information to the event representations, including absolute intensities (rather than intensity changes). As such, they can enhance the application results beyond what either imaging modality can achieve in isolation \cite{gallego_event-based_2022}. Eventually, there will be immense utility in an all-in-one adaptive streaming protocol for \textit{both} event-based and frame-based data. The ongoing MoQ efforts in frame-based streaming, then, can be integrated directly alongside a new event-based format.

\section{Implementation and Future Work}

At the time of writing, we are unaware of any open-source implementation of MoQ Transport that fully incorporates subgroups and delivery timeouts according to Draft 06 or later. In particular, these mechanisms are not yet available in the Rust-based \texttt{moq-transport} package \cite{english_englishmmoq-rs_2024} (commit ID fefb38f). Since our event video codec and client-side code were developed in Rust, we will focus on contributing to this package until it follows the latest draft. Then, we may evaluate our proposed event streaming format on the dataset from our prior work.

We believe that our effort into event-based streaming fills a major gap in the existing literature for computer vision with event cameras. As GPUs and vision applications get faster and event camera resolutions increase, it is increasingly necessary to have mechanisms for robust, low-latency event streaming. As JPEG XE moves towards a lossless compression standard in the coming years, we can transparently apply the codec to our proposed subgroup partitions. Loss, then, can be determined directly by the network conditions and application needs, rather than by a preset bitrate ladder at the camera source.

If there is wider interest in event-based streaming, we will propose our work as an MoQ Streaming Format. We expect that it will be useful to develop this format alongside MoQ Transport, which aims to be generic and handle arbitrary data payloads. Currently, the proposed MoQ Streaming Formats target traditional audio and video, chat messages, and server timestamp measurements. Our lossy protocol is complementary to these existing formats, opening the door to new optimizations for this unique data.

\section{Conclusion}

This work elucidates the need for rate-adaptive streaming protocols if event-based camera sensors are ever to gain traction in real-world systems. We analyzed the technical weaknesses of existing methods, including our own prior work, and identified how new constructs within the MoQ Transport draft may be leveraged for low-latency event streaming. Our proposed subgroup partitioning scheme and timeout-based rate adaptation set the stage for a new streaming format. The ongoing development of this format may inform future additions to the MoQ Transport protocol.

\bibliographystyle{ACM-Reference-Format}
\bibliography{acmart,references,references2}

\end{document}